\documentclass[prc,twocolumn,showpacs,showkeys,preprintnumbers,
nofootinbib,superscriptaddress,amsmath,amssymb]{revtex4}
\usepackage{graphics}
\usepackage{amsmath}
\usepackage{amssymb}
\newcommand{\etal}{\textit{et al.}}
\begin{document}
\title{Hints on the quadrupole deformation of the $\Delta$(1232)}
\author{C.~\surname{Fern\'andez-Ram\'{\i}rez}}
\email{cesar@nuc2.fis.ucm.es}
\affiliation{Instituto de Estructura de la Materia,
CSIC. Serrano 123, E-28006, Madrid. Spain.}
\affiliation{Departamento de F\'{\i}sica At\'omica, Molecular y Nuclear.
Universidad de Sevilla. Apdo. 1065, E-41080, Sevilla. Spain.}
\author{E.~\surname{Moya de Guerra}}
\affiliation{Instituto de Estructura de la Materia,
CSIC. Serrano 123, E-28006, Madrid. Spain.}
\affiliation{Departamento de F\'{\i}sica 
At\'omica, Molecular y Nuclear. Facultad de 
Ciencias F\'{\i}sicas. Universidad Complutense de Madrid. 
Avda. Complutense s/n, E-28040, Madrid. Spain.}
\author{J.M.~\surname{Ud\'{\i}as}}
\affiliation{Departamento de F\'{\i}sica 
At\'omica, Molecular y Nuclear. Facultad de 
Ciencias F\'{\i}sicas. Universidad Complutense de Madrid. 
Avda. Complutense s/n, E-28040, Madrid. Spain.}
\date{\today}
\begin{abstract}
The E2/M1 ratio (EMR) of the $\Delta$(1232) is 
extracted from the world data in pion photoproduction by means of
an Effective Lagrangian Approach (ELA).
This quantity has been derived within a 
crossing symmetric, gauge invariant, and chiral symmetric 
Lagrangian model which also contains a consistent modern treatment of the 
$\Delta$(1232) resonance.
The \textit{bare} s-channel $\Delta$(1232) 
contribution is well isolated and
Final State Interactions (FSI) are effectively taken into account 
fulfilling Watson's theorem.
The obtained EMR value,
EMR$=(-1.30\pm0.52)$\%,  
is in good agreement with the 
latest lattice QCD calculations [Phys. Rev. Lett. 94, 021601 (2005)]
and disagrees with results of current quark model calculations.
\end{abstract}
\maketitle

From general symmetry principles, 
the emission of a photon by a spin-3/2 system that becomes
spin-1/2, involves transverse electric quadrupole (E2) and 
magnetic dipole (M1) multipolarities. 
Likewise, this is the case of
the absorption of a real photon by a spin-1/2 to reach spin-3/2.
In the absence of knowledge of the internal structure of the system, 
an estimate of the ratio between the two multipolarities can be made by 
resorting to Weisskopf \cite{Weisskopf} units for multipole 
strengths in nuclear systems.
For the excitation of a nucleon into a $\Delta$(1232) 
($\gamma +  N \to \Delta$)
this estimate gives
\begin{equation}
R_W=\sqrt{\left( \frac{S_{\text{E2}}}{S_{\text{M1}}} \right)}=
1.07 \cdot 10^{-3} R_0^2 \left( M_\Delta - M_N \right) \: ,
\end{equation}
with the nucleon radius $R_0$ in fm and the mass difference in MeV.
In what follows we refer to this value as the Weisskopf ratio ($R_W$).
Taking a radius $R_0=0.875$ \cite{PDG2004}
and a mass difference $\left( M_\Delta - M_N \right)\simeq 270$ MeV 
one gets $R_W \simeq 0.22$.

Within the quark model, a single quark spin flip is the standard picture
for the photoexcitation of the nucleon into a $\Delta$, assuming
spherically symmetric ($L=0$) radial wave functions of both parent 
and daughter. Under these premises, an E2 transition cannot take place,
as it was first noticed by Becchi and Morpurgo in their
1965 paper \cite{Becchi}, where they concluded that a value of the 
$\text{E2}/\text{M1}$ ratio (EMR) 
much smaller than $R_W$ should be considered as a test
of the model. As early as 1963 values of EMR small but different from zero
were reported in the literature \cite{Gourdin}
which was supported by further experiments 
later on \cite{Davidson,Blanpied,Mainz}.
A non-vanishing E2 multipolarity evokes 
a deformed nucleon picture \cite{Glashow}.
In an extreme rotational model approximation the nucleon could be considered
as the head of a $K^\pi=\frac{1}{2}^+$ rotational band 
($\frac{1}{2}^+,\frac{3}{2}^+,\frac{5}{2}^+, \dots$), 
in analogy to rotational nuclear bands.
In this picture the electromagnetic current and
multipoles for the transition between the members of the band
can be parametrized in terms of intrinsic single particle and collective
multipoles \cite{EMoya}.
In particular, the E2 multipole for the transition ($\gamma +  N \to \Delta$)
would be given in terms of the intrinsic quadrupole moment ($Q_0$) 
by the relation \cite{EMoya,Bohr}
\begin{equation}
\mathcal{M}\left( \text{E2} \right)=
< \frac{1}{2} \frac{1}{2} 2 0 | \frac{3}{2} \frac{1}{2} > 
\sqrt{\frac{5}{8\pi}} Q_0 = 0.282 Q_0 \: .
\end{equation}
In turn, $Q_0$ would be related to the spectroscopic 
quadrupole moment of the $\Delta$ by 
\begin{equation}
Q_0=-5Q_\Delta \: .
\end{equation}

Hence, the relationship between the static $\Delta$(1232) 
quadrupole moment and the E2 multipole for the $N \to \Delta$ transition is 
\begin{equation}
\mathcal{M}\left(\text{E2} , N \to \Delta \right)
=-\frac{5}{\sqrt{4\pi}} Q_\Delta \:.
\end{equation}

Within this picture, a negative (positive) static quadrupole moment
implies a prolate (oblate) intrinsic deformation,
which is not always well stated in the literature.

Over the last few years much effort has been invested in 
the determination of 
quadrupole deformation in 
the nucleon \cite{Bernstein,Krusche}.
Because the spin of the nucleon is 1/2, a possible intrinsic
quadrupole deformation is not directly observable and 
its study requires research on its lowest-lying excitation 
-- $\Delta$(1232) --
and its decay through pion emission. Hints on the possible deformation
will be deduced via the EMR.  
In the context of the  quark model,
De R\'ujula, Georgi, and Glashow \cite{DeRujula} 
were the first to suggest a tensor force arising from one-gluon 
exchange and leading to d-state admixtures. 
On the other hand Buchmann and collaborators \cite{Buchmann} 
pointed out that a non-zero E2 transition could be due to one-gluon or 
meson-exchange currents.
While debate on the physical interpretation of the EMR 
may still be far from closed, a more precise determination 
of the EMR value is both possible and mandatory.

Extensive experimental programs have been
developed at Brookhaven \cite{Blanpied}
and Mainz \cite{Mainz}, that have resulted in an 
improvement in the quantity and quality of the pion photoproduction
data \cite{Drechsel}.
However, in order to extract the EMR from 
experiment, a realistic model of the reaction must be employed 
that takes into account the Final State Interaction (FSI)
of the outgoing pion
as well as  the relevant symmetries. 
Only then can the ratio deduced from the
experimental data be compared to the predictions of nucleonic models 
| namely, quark models \cite{Becchi,Buchmann}, 
skyrme models \cite{Wirzba}, and
lattice QCD \cite{Leinweber,Alexandrou}.
Theoretical
interest in this topic has been strongly renewed and
either new or well-known approaches have been (re)investigated with
the latest theoretical advances such as 
new dynamical models \cite{Sato,Fuda,Tjon}
and non-pathological spin-3/2 treatments \cite{Tjon,cefera}.
A complete account of the experimental and theoretical work done on
this topic goes well beyond the scope of this paper. For a review
of the subject we refer the reader to Ref. \cite{Krusche}.

A key point in the extraction of the EMR is the reaction model
used for the analysis of data.
Reaction models have to be developed 
carefully 
in order to consider the underlying physics and to
minimize model dependencies as well as 
theoretical uncertainties.
Ambiguities in the contribution of the background
terms, unitarization, 
or even formal elements 
(such as the recently improved spin-3/2 description
or the crossing symmetry)
can spoil the determination of the parameters of the 
resonances.
This is so even for a well isolated
resonance as is the $\Delta$(1232).
A determination of the $\Delta$(1232) parameters requires one to
study the photoproduction reaction not only 
in the first resonance region, as commonly has been done,
but in further kinematical regions
in order to keep under control the high energy behavior of the
resonance contribution.
For example, in a Breit-Wigner model, 
the inclusion of Regge poles, which take into account 
heavy meson exchanges, does affect the determination of the 
$\Delta$(1232) coupling constants because of the modification
of the tail of the resonance \cite{Aznauryan}.

\begin{figure}
\begin{center}
\scalebox{0.65}{\includegraphics{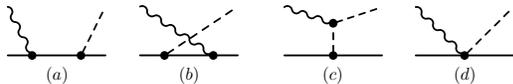}}
\end{center}
\caption{Feynman diagrams for Born terms: ($a$) s-channel, 
($b$) u-channel, ($c$) t-channel, 
and ($d$) Kroll-Rudermann.} \label{fig:diag1}
\end{figure}

From the theoretical point of view,
the Effective Lagrangian Approach (ELA)
is a very suitable 
and appealing method to study pion photoproduction 
and nucleon excitations.
It is also a reliable, accurate, and formally well established
approach in the nucleon mass 
region.

\begin{figure}
\begin{center}
\scalebox{0.65}{\includegraphics{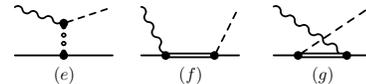}}
\end{center}
\caption{Feynman diagrams for vector meson exchange ($e$) and 
resonance excitations: ($f$) s-channel and ($g$) 
u-channel.} \label{fig:diag2}
\end{figure}

In this Rapid Communication 
we employ a realistic model 
for pion photoproduction 
on free nucleons from threshold up to 1 GeV
based on the ELA 
that we have recently elaborated.
Details on the model will be published somewhere else
and can be found in \cite{cefera}.
In what follows we provide a brief description of the model.
In addition to Born (Fig. \ref{fig:diag1})
and vector meson exchange terms 
($\rho$ and $\omega$, diagram ($e$) in Fig. \ref{fig:diag2}), 
the model includes 
all the four star resonances in Particle Data Group (PDG)
\cite{PDG2004} up to 1.7 GeV mass
and up to spin-3/2:
$\Delta$(1232), N(1440), N(1520), $\Delta$(1620), N(1650), 
and $\Delta$(1700) --- diagrams ($f$) and ($g$) in Fig. \ref{fig:diag2}.
The main advantages of our
model compared to previous ones \cite{pions} 
resides on the treatment of resonances. 
In particular, we avoid some pathologies in the Lagrangians 
of the spin-3/2 resonances (such as $\Delta$(1232)),
present in previous models,
implementing a modern approach 
due to Pascalutsa \cite{Pascalutsa}.
Under this approach the (spin-3/2 resonance)-nucleon-pion and 
the (spin 3/2 resonance)-nucleon-photon
vertices have to fulfill the condition
$q_\alpha {\mathcal O}^{\alpha ...}=0$ where $q$ is the four-momentum 
of the spin-3/2 particle, $\alpha$ the vertex index which couples 
to the spin-3/2 field, and the dots stand for other possible 
indices. In particular, we write the simplest interacting 
(spin-3/2 resonance)-nucleon-pion 
Lagrangian as \cite{Pascalutsa}
\begin{equation}
{\mathcal L}_{int}=-\frac{h}{f_\pi M^*} \bar{N} \epsilon_{\mu 
\nu \lambda \beta} \gamma^\beta \gamma^5 \left( \partial^\mu 
N^{*\nu}_j \right) \left( \partial^\lambda \pi_j  \right) + 
HC \: ,\label{GIcoupling}
\end{equation}
where $HC$ stands for hermitian conjugate,
$h$ is the strong coupling constant, 
$f_\pi=92.3$ MeV is the leptonic
decay constant of the pion, $M^*$ the mass of the resonance, and
$\pi_j$, $N$, and $N^{*\nu}_j$, the pion, nucleon, and spin-3/2
fields respectively.

The model also displays 
chiral symmetry, gauge invariance,
and crossing symmetry.
The dressing of the resonances \cite{Leupold} is considered by means of a 
phenomenological width which takes into account 
decays into one $\pi$, one $\eta$, and two $\pi$.
The width is built in order to fulfill crossing symmetry and contributes
to both s- and u- channels of the resonances.
In order to regularize the high energy behavior of the model we include
a crossing symmetric and gauge invariant form factor for Born and
vector meson exchange terms 
\cite{Davidson01-1}, as well as
form factors in the resonance contributions
consistent with the phenomenological widths.
We assume that the FSI factorizes and can be 
included through the distortion of the 
$\pi N$ final state wave function. 
Factorization of FSI
has been successfully 
applied to electron scattering knock-out reactions \cite{Udias}.

A detailed calculation of the distortion would require one to  
calculate higher order pion loops or to develop a 
phenomenological potential FSI model.
The first approach is overwhelmingly complex
and the second would introduce additional model-dependencies,
which are to be avoided in the present analysis, 
in as much as we are 
concerned here with the bare properties 
of the $\Delta$(1232).
We rather include FSI in a phenomenological way by
adding a phase $\delta_{\text{FSI}}$ to the electromagnetic multipoles.
We determine this phase so that the total phase of the electromagnetic
multipole is identical to the one of the  
energy dependent solution of SAID \cite{SAIDdata}.
In this way
Watson's theorem \cite{Watson} is fulfilled below the two pion threshold
and we are able to disentangle 
the electromagnetic vertex from FSI effects.

In order to obtain a reliable set of
electromagnetic coupling constants of the nucleon resonances
we have fitted the experimental electromagnetic multipoles
using modern minimization techniques
based upon genetic algorithms.
We have obtained
different fits which are compared in Ref. \cite{cefera}. 
In this Rapid Communication 
we focus on two fits obtained including FSI 
and using two different prescriptions 
for the determination of the masses of the resonances.
The first one uses the set of masses
and widths provided by Vrana, Dytman, and Lee \cite{Vrana}, 
and the second one uses a set
established by means of a speed plot calculation from
the current solution of the SAID $\pi N$ partial wave analysis 
\cite{SAIDdata}.

\begin{figure}
\begin{center}
\rotatebox{0}{\scalebox{0.4}[0.4]{\includegraphics{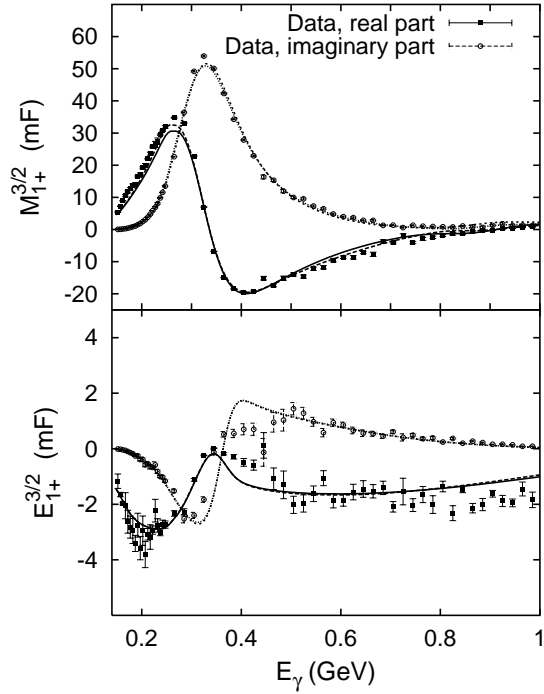}}}
\end{center}
\caption{Comparison among the fits from sets \#1 
(solid: real part, short-dashed: imaginary part) 
and \#2 (dashed: real part, dotted: imaginary part) and the
SAID energy independent solution (data) for
$M_{1+}^{3/2}$ and $E_{1+}^{3/2}$ electromagnetic multipoles \cite{SAIDdata}.
A detailed discussion on electromagnetic multipoles can be found
in Ref. \cite{cefera}.} 
\label{fig:delta}
\end{figure}

In Fig. \ref{fig:delta} we show our 
fits to $M_{1+}^{3/2}$ and $E_{1+}^{3/2}$ multipoles
for both sets of parameters.

Caution must be taken with the various definitions of EMR
employed in the literature.
We should distinguish between 
the intrinsic (or \textit{bare} EMR of the $\Delta$(1232) and
the directly measured
value which is often called \textit{physical} or \textit{dressed} 
EMR value \cite{Sato,Tjon}
and which is obtained as the ratio between the 
imaginary parts of $E_{1+}^{3/2}$
and $M_{1+}^{3/2}$ at the $E_\gamma$ value at which 
$\text{Re}M_{1+}^{3/2}=0=\text{Re}E_{1+}^{3/2}$.
Since all the reaction models are fitted to the experimental electromagnetic
multipoles, they generally reproduce the physical EMR value. As seen
in Fig. \ref{fig:delta} this is also the case in our model, where
we get 
\begin{equation}
\frac{\text{Im}E_{1+}^{3/2}}{\text{Im} M_{1+}^{3/2}}
=\left( -3.9 \pm 1.1 \right) \% 
 \end{equation}
for $328 \text{ MeV} \leq E_\gamma \leq 343 \text{ MeV}$.
This value compares well
with the value obtained by LEGS Collaboration in \cite{Blanpied}, 
$\left( -3.07\pm 0.26 \rm (stat.+syst.) \pm 0.24 \rm (model) \right) \%$, 
and is somewhat higher than the PDG value
$\left( -2.5\pm 0.5 \right) \%$.

However, this measured EMR value is not easily 
computed with the theoretical models
 of the nucleon and its resonances. 
Instead, in order to compare to models of nucleonic structure, 
it is better to extract the \textit{bare} EMR value
of $\Delta$(1232) which is defined as:
\begin{equation}
\text{EMR}=\frac{G_E^\Delta}{G_M^\Delta}\times 100\% \: , \label{eq:EMR}
\end{equation} 
This depends only on the intrinsic characteristics 
of the $\Delta$(1232) and can thus
be compared directly to predictions from  nucleonic models. It is not, however,
directly measurable but must be inferred (in a model dependent way) 
from reaction models, precisely what we aim in this Rapid Communication.

\begin{table}
\caption{Intrinsic (or bare) EMR (from eq. (\ref{eq:EMR})) 
and parameters of $\Delta$(1232)
for the two fits considered.
$M_\Delta$ is the mass,
$A_{1/2}^\Delta$ and $A_{3/2}^\Delta$ the helicity amplitudes,
$G_E^\Delta$ the electric form factor, and
$G_M^\Delta$ the magnetic form factor.
Masses and widths for set \#1 have been taken from Ref. \cite{Vrana}
and for set \#2 they have been calculated using the 
speed plot technique \cite{cefera}.} \label{tab:ratios}
\begin{ruledtabular}
\begin{tabular}{lcc}

 & Set \#1 & Set \#2 \\

\hline
$M_\Delta$ (MeV)      & $1215\pm2$&$ 1209\pm2$ \\
$A_{1/2}^\Delta$ (GeV$^{-1/2}$) &$-0.123\pm0.003$&$-0.123\pm0.003$ \\
$A_{3/2}^\Delta$ (GeV$^{-1/2}$) &$-0.225\pm0.005$&$-0.224\pm0.004$ \\
$G_E^\Delta$            &$-0.076\pm0.042$&$-0.071\pm0.042$ \\
$G_M^\Delta$            &$ 5.650\pm0.070$&$ 5.701\pm0.071$ \\
EMR    &$\left( -1.35\pm0.74 \right)\%$&$\left( -1.24\pm0.74\right)\%$ \\
\end{tabular}
\end{ruledtabular}
\end{table}

The connection between both definitions of EMR values 
is straightforward when FSI are neglected as can be found in the paper
by Jones and Scadron \cite{Jones}. In our formalism, both values 
can be connected from the
definitions of the electromagnetic multipoles \cite{cefera} and their
connection to the  $\gamma +  N \to \Delta$ transition Lagrangian
\begin{equation}
{\mathcal L}_{em}=\frac{3e}{2M M^+} \bar{N} 
\left[ ig_1 \tilde{F}_{\mu \nu}+g_2 \gamma^5 F_{\mu \nu} 
\right] \partial^\mu N^{*\nu}_3 + HC \: ,
\end{equation}
where $M^+=M+M_\Delta$,
$F_{\mu \nu}$ is the electromagnetic field,
$\tilde{F}_{\mu \nu}=\frac{1}{2}
\epsilon_{\mu \nu \alpha \beta} F^{\alpha \beta}$,
and $g_1$ and $g_2$ are the coupling constants that can be related to 
the electric and magnetic form factors through
$G_E^\Delta=-\frac{1}{2}\frac{M_\Delta - M}{M^+}g_2$ and 
$G_M^\Delta= g_1 + \frac{1}{2}\frac{M_\Delta - M}{M^+}g_2$ \cite{Pasc99}.
In our calculation, the numerical differencies between the dressed and the
bare EMR values are attributed to FSI.

In Table \ref{tab:ratios} we quote
our extracted bare EMR values obtained from Eq. (\ref{eq:EMR}) together
with the mass, helicity amplitudes, and 
electromagnetic form factors at the photon point of the $\Delta$(1232).

In our calculations we have considered the pole mass of the resonance
instead  of the Breit-Wigner mass \cite{Sato,Fuda,Tjon}.
One must be aware of the fact that electromagnetic coupling 
constants are very sensitive 
to the mass and that the width of the $\Delta$(1232) 
and the multipoles vary rapidly 
in the region around the peak of the $\Delta$(1232).
Thus, a variation in the mass 
of the resonance affects the determination of the EMR value.
This is also seen in Table \ref{tab:ratios}.
Out of the two results given in Table \ref{tab:ratios} 
we adopt as our final result the average value 
for the bare EMR$=\left( -1.30\pm0.52 \right) \%$.

\begin{table*}
\caption{Comparison of EMR values from nucleonic models and 
EMR values extracted from data predicted 
through several reaction models (see text).} \label{tab:models}
\begin{ruledtabular}
\begin{tabular}{lcc}
Physical EMR, experiments & EMR& Ref.\\ 
\hline
Particle Data Group & $\left(-2.5 \pm 0.5 \right) \%$ & \cite{PDG2004} \\
LEGS Collaboration  & 
$\left(-3.07 \pm 0.26 {(\rm stat.+syst.)}\pm 0.24 {\rm (model)}\right) \%$ 
& \cite{Blanpied} \\
\hline
Physical EMR, reaction models & & \\ 
\hline
Sato and Lee         &$-2.7$\%        & \cite{Sato} \\
Fuda and Alharbi     &$-2.09$\%       & \cite{Fuda} \\
Pascalutsa and Tjon  &$\left( -2.4\pm0.1 \right)$\%   & \cite{Tjon} \\
\textbf{Present work (average)}&$\left( \mathbf{-3.9 \pm 1.1} \right) \%$
& \\

\hline
Extractions of bare EMR, reaction models &  & \\ 

\hline
Sato and Lee         &$-1.3$\%        & \cite{Sato} \\
Pascalutsa and Tjon  &$\left( 3.8\pm1.6 \right)$\%   & \cite{Tjon} \\
\textbf{Present work (average)}&$\left( \mathbf{-1.30\pm0.52} \right) \%$
& \\
\hline

Bare EMR, predictions from nucleonic models & &  \\

\hline
Non-relativistic quark model& 0\% & \cite{Becchi}\\
Constituent quark model &$-3.5$\% & \cite{Buchmann} \\
Skyrme model& $\left( -3.5\pm1.5 \right) $\%& \cite{Wirzba}\\
Lattice QCD (Leinweber \etal)& $\left( 3 \pm 8 \right)$\%
& \cite{Leinweber}\\
Lattice QCD (Alexandrou \etal)& &\cite{Alexandrou} \\
 ($Q^2=0.1$ GeV$^2$, $m_\pi = 0$)
&$\left( -1.93 \pm 0.94 \right)$\%& \\
 ($Q^2=0.1$ GeV$^2$, $m_\pi = 370$ MeV)
&$\left( -1.40 \pm 0.60 \right)$\%& \\

\end{tabular}
\end{ruledtabular}
\end{table*}

In Table \ref{tab:models} we compare our average EMR values 
(bare and dressed) to
the ones extracted by other authors using other models for pion
photoproduction, as well as to 
predictions of nucleonic models.
Our bare result is similar to that from Ref. \cite{Sato}.
However, it disagrees with the bare value derived with 
the dynamical model of 
Pascalutsa and Tjon \cite{Tjon}, where a positive deformation
of the $\Delta$(1232) (EMR$=\left( 3.8\pm1.6 \right)\%$) is inferred.
We compare to their model because, together with the one 
we employ in this work,
they were the only available models that include 
non-pathological $\Delta$(1232)
Lagrangians. The discrepancy 
is not so worrysome if we recall that 
dynamical models have ambiguities in the determination 
of the bare value of EMR \cite{Wilhelm} that is highly
model dependent as it stems 
from the comparison among different dynamical models, 
namely Refs. \cite{Sato,Fuda,Tjon}.
More recently \cite{Pasc05} the dependence of the 
effective chiral perturbation theory on the small expansion parameters
was fully exploited to reconcile the (bare) lattice QCD calculations
with the physical EMR values.

In conclusion, the bare
EMR value derived from the multipole experimental data with our realistic
ELA model is compatible with some of the predictions of the nucleonic models.
In particular it agrees very well with the latest 
lattice QCD calculations \cite{Alexandrou} 
and suggests the need for further improvements in quark models.
The comparison of our extracted EMR value to $R_W$ is indicative
of a small oblate deformation of the $\Delta$(1232).
In our work we show that an ELA which
takes into account FSI is also able to 
reconcile the physical EMR value with the lattice QCD calculations
prediction for EMR. 
We consider that our picture and that of Ref. \cite{Pasc05}
are complementary. Thus, both pictures will help to understand
the issue of the $\Delta$(1232) deformation as well as the
properties of other resonances.

\begin{acknowledgments}
C.F.-R. work is being developed under Spanish 
Government grant UAC2002-0009.
This work has been supported in part under contracts of 
Ministerio de Educaci\'on y Ciencia (Spain) 
BFM2002-03562, BFM2003-04147-C02-01, and FIS2005-00640.
\end{acknowledgments}

\end{document}